\documentclass[a4paper,12pt]{article}
\usepackage[T2A]{fontenc}
\usepackage[utf8]{inputenc}
\usepackage{amsmath}\numberwithin{equation}{section}
\usepackage{amsfonts}
\usepackage{mathrsfs}
\usepackage{amssymb}
\usepackage{cite}
\usepackage{color}

\newcommand{\be}{\begin{equation}}
\newcommand{\ee}{\end{equation}}
\newcommand{\bea}{\begin{eqnarray}}
\newcommand{\eea}{\end{eqnarray}}

\newcommand{\p}[1]{(\ref{#1})}

\begin{document}



\begin{center}

{\LARGE\bf On the off-shell superfield Lagrangian formulation }

\vspace{0.15cm}

{\LARGE\bf of $4D$, $\mathcal{N}{=}\,1$ supersymmetric infinite spin theory}

\vspace{1.1cm}

{\large\bf I.L.\,Buchbinder$^{1,2,3,4}$\!\!,\ \ \
S.A.\,Fedoruk$^4$\!\!,\ \ \ A.P.\,Isaev$^{4,5}$\!\!,\ \ \
V.A.\,Krykhtin$^{1,2}$}

\vspace{1.2cm}

\ $^1${\it Center of Theoretical Physics, Tomsk State Pedagogical University, \\
634041 Tomsk, Russia}, \\
{\tt joseph@tspu.edu.ru, krykhtin@tspu.edu.ru}

\vskip 0.5cm

\ $^2${\it National Research Tomsk State  University, \\
634050 Tomsk, Russia}

\vskip 0.5cm

\ $^3${\it Lab of Theor. Cosmology, International Center of Gravity
    and Cosmos,\\
    Tomsk State University of Control Systems and Radioelectronics,\\
    (TUSUR), 634050, Tomsk, Russia}

\vskip 0.5cm

\ $^4${\it Bogoliubov Laboratory of Theoretical Physics, Joint Institute for Nuclear Research, \\
141980 Dubna, Moscow Region, Russia}, \\
{\tt fedoruk@theor.jinr.ru, isaevap@theor.jinr.ru}

\vskip 0.5cm

\ $^5${\it St.Petersburg Department of Steklov Mathematical Institute of RAS, \\
Fontanka 27, 191023 St. Petersburg, Russia}

\end{center}

\vspace{0.7cm}

\begin{abstract}
We develop a complete off-shell Lagrangian description of the free
$4D, {\cal N}=1$ supersymmetric theory of infinite spin. Bosonic and
fermionic fields are formulated in terms of spin-tensor
fields with dotted and undotted indices. The corresponding
Lagrangians for bosonic and fermionic infinite spin fields
entering into the on-shell supersymmetric model are derived within
the BRST method. Lagrangian for this supersymmetric model is written
in terms of the complex infinite spin bosonic field and infinite
spin fermionic Weyl field subject to supersymmetry transformations.
The fields involved into the on-shell supersymmetric Lagrangian can
be considered as components of six infinite spin chiral and
antichiral multiplets. These multiplets are extended to the
corresponding infinite spin chiral and antichiral superfields so
that two chiral and antichiral superfields contain among the
components the basic fields of an infinite spin supermultiplet and
extra four chiral and antichiral superfields containing only the
auxiliary fields needed for the Lagrangian formulation. The
superfield Lagrangian is constructed in terms of these six chiral
and antichiral supefields, and we show that the component form of
this superfield Lagrangian exactly coincides with the previously
found component supersymmetric Lagrangian after eliminating the
component fields added to construct (anti)chiral superfields.

\end{abstract}

\vspace{0.4cm}



\setcounter{footnote}{0}
\setcounter{equation}{0}


\section{Introduction}
The study of various aspects of massless fields corresponding to
the infinite spin irreducible representations of the Poincar\'{e} group
\cite{Wigner39,Wigner47,BargWigner} attracts attention due to
their remarkable relations with higher spin theory and string theory
\cite{Iverson-Mack, Brink:2002zx, Schuster:2014hca,
Rivelles:2014fsa, Metsaev:2016lhs, Metsaev:2018lth,
Khabarov:2017lth, Najafizadeh:2015uxa, Metsaev:2017ytk,
Metsaev:2017, Najafizadeh:2017tin, Mourad:2005rt, Metsaev:2017cuz,
Bekaert:2017xin, Bekaert:2017khg, Fedoruk, Bengtsson:2013vra,
Alkalaev:2017hvj, ACG18, Metsaev18a, BFI, Metsaev19,
Buchbinder:2018yoo,Buchbinder:2020nxn,
Zinoviev:2017rnj,Buchbinder:2019iwi,Buchbinder:2019esz,Buchbinder:2019olk,
Khabarov:2020glf,Najafizadeh:2019mun,Najafizadeh:2021dsm}. One of
such interesting aspects is the problem of constructing a
supersymmetric infinite spin theory and the corresponding Lagrangian
formulation
\cite{Zinoviev:2017rnj,Buchbinder:2019iwi,Buchbinder:2019esz,Buchbinder:2019olk,
Khabarov:2020glf,Najafizadeh:2019mun,Najafizadeh:2021dsm}.

As is well known, $\mathcal{N}=1$ supersymmetric field theories can
be defined in two ways. First, the theory is formulated in terms of
ordinary bosonic and fermion fields subject to supersymmetry
transformations. These fields are called components. In this
formulation, supersymmetry is nonmanifest; moreover, if the
Lagrangian does not contain special non-dynamical auxiliary fields,
the supersymmetry algebra is only on-shell closed. To avoid
ambiguities, it is worth emphasizing that the term of the auxiliary
fields is used in supersymmetric higher spin theories in two
completely different meanings. Firstly, these are the fields that
ensure the closure of the supersymmetry algebra. Secondly, this term
is used for non-dynamic fields that provide  the Lagrangian
formulation in higher spin field theory. Since infinite spin
theories are similar in some respects to higher spin theories, they
can also be characterized by the auxiliary fields irrespective of
supersymmetry. Another way to formulate supersymmetric theories is
based on the use of unconstrained superfields (see, e.g.,
\cite{Buchbinder:1998qv}), which automatically provide manifest
supersymmetry and a closed superalgebra\footnote{Certainly, the
constrained superfields or light-cone superfields are useful as
well, although they can lead to violation of manifest symmetries. }.

In the present paper, we solve the problem of constructing the
manifest off-shell Lagrangian formulation of $4D, \mathcal{N}=1$
supersymmetric free infinite spin fields. Unlike all previous
papers, we give a complete superfield description and derive the
Lagrangian in terms of six scalar chiral and antichiral superfields.
Two of these superfields contain among the components the basic
fields corresponding to an infinite spin supermultiplet
\cite{Buchbinder:2019esz} and the other superfields are exclusively
auxiliary in sense of the higher spin theories. It should be noted that
we work within the BRST approach (see application of this approach to
infinite spin theories in
\cite{Buchbinder:2018yoo,Buchbinder:2020nxn}), where all the fields
are understood as components of the Fock space vectors. Therefore,
more precisely, when we talk about superfields, we mean the
corresponding vectors of the Fock space with superfield components.

The description of $4D, \mathcal{N}=1$ supersymmetric infinite spin
theories was studied in the component approach in recent papers
\cite{Khabarov:2020glf,Najafizadeh:2019mun,Najafizadeh:2021dsm},
where supersymmetry transformations were found for one complex
scalar field (vector of Fock space in our formulation) and the Dirac
field (the Fock space vector in our formulation), in addition, in
the paper \cite{Najafizadeh:2021dsm} off-shell  supersymmetry
transformations (up to gauge transformations) were given for
component fields. Unlike the previous works on supersymmetric
infinite spin theories we use Lagrangians for bosonic and fermionic
fields which were obtained using the BRST method
\cite{Buchbinder:2018yoo,Buchbinder:2020nxn}. It was noted in
\cite{Buchbinder:2020nxn} that Lagrangian for fermionic infinite
spin obtained there contains twice as many degrees of freedom as
Metsaev's Lagrangian \cite{Metsaev:2017ytk}. Therefore we split
Lagrangian obtained in  \cite{Buchbinder:2020nxn} in terms of Dirac
spinors into two parts which are written  in terms of Weyl spinors,
each of which will describe the fermionic field of a infinite spin.
Thus we consider the Lagrangian system of one dynamical complex
scalar infinite spin field and one dynamical Weyl infinite spin
field (both are the vectors of the Fock space in our formulation).
On the basis of these fields, by adding some new additional fields
needed for composing (anti)chiral superfields, we construct a
superfield Lagrangian formulation for supersymmetric infinite spin
theory. Our final result is the first completely superfield
Lagrangian formulation for the $4D, \mathcal{N}=1$ infinite spin
field theory.

The paper is organized as follows. In the next section, we recall the
Lagrangian construction for fermionic and bosonic continuos spin
fields constructed in terms of spin-tensor fields with dotted and
undotted indices within the BRST approach suggested in
\cite{Buchbinder:2018yoo,Buchbinder:2020nxn}. In
section~\ref{sec:SField}, we first consider the sum of Lagrangians of
one complex bosonic field and one fermionic Weyl field which are
the vectors of the Fock space and find for such system the supersymmetry
transformations in component form. After this, we discuss
construction of superfields from these component fields by
adding some new auxiliary fields needed for composing (anti)chiral
superfields. In section~\ref{sec:SFieldLagr}, by using  the resulting superfields,
we construct a superfield Lagrangian and present
gauge transformation for the superfields. Then we show that after elimination of the redundant fields the
superfield Lagrangian indeed reproduces the component Lagrangian given
section~\ref{sec:SField}. In Summary, we discuss the results  obtained in the paper.

\section{BRST Lagrangian formulation of infinite spin fields}

The description of fields of infinite spin was presented in
\cite{Buchbinder:2018yoo} for the case of integer spins and in
\cite{Buchbinder:2020nxn} for half-integer spins. Here we summarize
the results obtained there for the subsequent formulation of the
supersymmetric generalization.

\subsection{Generalized infinite spin Fock space}

Physical states are defined in the generalized Fock space.
For this reason, we introduce the operators
\begin{equation}
\label{a-oper}
a_\alpha\,,\quad b^\alpha\,,
\end{equation}
which are Weyl spinors and satisfy the algebra
\begin{equation}
\label{a-com}
[a_\alpha,b^\beta]=\delta_\alpha^\beta\,.
\end{equation}
Hermitian conjugation yield the operators
\begin{equation}
\label{c-oper}
\bar{a}_{\dot\alpha}=(a_\alpha)^\dagger\,,\quad
\bar{b}^{\dot\alpha}= (b^\alpha)^\dagger\,,
\end{equation}
with the commutation relation
\begin{equation}
\label{c-com}
[\bar{b}^{\dot\alpha},\bar{a}_{\dot\beta}]=\delta^{\dot\alpha}_{\dot\beta}\,.
\end{equation}
Below we use the notation
\begin{equation}
\label{not-oper}
a_{\alpha(s)}:=a_{\alpha_1}\ldots a_{\alpha_s}\,,\quad \bar a_{\dot\alpha(s)}:=\bar a_{\dot\alpha_1}\ldots \bar a_{\dot\alpha_s}\,,\quad
b^{\alpha(s)}:=b^{\alpha_1}\ldots b^{\alpha_s}\,,\quad \bar b^{\dot\alpha(s)}:=\bar b^{\dot\alpha_1}\ldots \bar b^{\dot\alpha_s}\,.
\end{equation}

Following \p{a-com} and \p{c-com}, we consider the operators
$a_\alpha$ and  $\bar{b}^{\dot\alpha}$ as annihilation operators and
define the ``vacuum'' state
\begin{equation}\label{vac}
|0\rangle\,, \qquad \langle0|=(|0\rangle)^\dagger\,,\qquad\langle0|0\rangle=1
\end{equation}
by the relations
\begin{equation}\label{caoperators}
a_\alpha|0\rangle=\bar{b}^{\dot\alpha}|0\rangle=0\,,
\qquad\langle0|\bar{a}_{\dot\alpha}=\langle0|b^\alpha=0
\,.
\end{equation}

The Fock space is formed by two types of vectors.
Vectors of the first type have the form
\begin{equation} \label{GFState-b}
|\varphi\rangle=
\sum_{s=0}^{\infty}|\varphi_{s}\rangle \,,\qquad
|\varphi_{s}\rangle:=\frac{1}{s!}\,\varphi_{\alpha(s)}{}^{\dot\beta(s)}(x)\ b^{\alpha(s)}\,\bar{a}_{\dot\beta(s)}|0\rangle \,.
\end{equation}
Then the  conjugate vector to (\ref{GFState-b}) is written as follows:
\begin{equation} \label{bGFState-b}
\langle\bar{\varphi}|=
\sum_{s=0}^{\infty}\langle\bar{\varphi}_s|\,,\qquad
\langle\bar{\varphi}_s|:= \frac{1}{s!}\,\langle 0|\,\bar{b}^{\dot\alpha(s)}\,a_{\beta(s)}\ \bar{\varphi}^{\beta(s)}{}_{\dot\alpha(s)}(x).
\end{equation}
Expansions \p{GFState-b} and \p{bGFState-b} contain an equal
number of operators with undotted and dotted indices. All component fields
$\varphi_{\alpha(s)}{}^{\dot\beta(s)}(x)$ and $\bar{\varphi}^{\beta(s)}{}_{\dot\alpha(s)}(x)$ have bosonic statistics,
i.e. these fields are $c$-number.

The Fock space, which we consider, also contains the vectors with external Dirac index $A=1,2,3,4$:
\begin{equation} \label{GFState}
|\Upsilon_A\rangle=
\sum_{s=0}^{\infty}|\Upsilon_{A\,,s}\rangle \,,\qquad
|\Upsilon_{A,s}\rangle:=\frac{1}{s!}\,\Upsilon_{A\,\alpha(s)}{}^{\dot\beta(s)}(x)\ b^{\alpha(s)}\,\bar{a}_{\dot\beta(s)}|0\rangle \,.
\end{equation}
The Dirac conjugate vector to (\ref{GFState}) is written as follows:
\begin{equation} \label{bGFState}
\langle\bar{\Upsilon}^A|=\left( |\Upsilon_B\rangle \right)^\dagger(\gamma_0)_B{}^A=
\sum_{s=0}^{\infty}\langle\bar{\Upsilon}^A_s|\,,\qquad
\langle\bar{\Upsilon}^A_s|:= \frac{1}{s!}\,\langle 0|\,\bar{b}^{\dot\alpha(s)}\,a_{\beta(s)}\ \bar{\Upsilon}^{A\,\beta(s)}{}_{\dot\alpha(s)}(x).
\end{equation}
Expansions \p{GFState} and \p{bGFState} contain an equal
number of operators with undotted and dotted indices, like
expressions \p{GFState-b} and \p{bGFState-b}.
However, the component fields $\Upsilon_{A\,\alpha(s)}{}^{\dot\beta(s)}(x)$ and $\bar{\Upsilon}^{A\,\beta(s)}{}_{\dot\alpha(s)}(x)$
are Grassmann variables and obey fermionic statistics.

It is natural that for the bosonic creation and annihilation operators the vectors \p{GFState-b}, \p{bGFState-b}
are bosonic while the vectors \p{GFState}, \p{bGFState} with external Dirac index~$A$ are fermionic.

We introduce the operators
\begin{equation} \label{l-def}
l_0:=\partial^2=\Box\,,\qquad  l_1:= i\,a^\alpha\bar{b}^{\dot\beta}\partial_{\alpha\dot\beta}\,,\qquad
l_1^+:=i\,b^\alpha\bar{a}^{\dot\beta}\partial_{\alpha\dot\beta}
\end{equation}
in our Fock space. The nonzero commutator of the above operators is
\begin{equation}\label{algebra}
[l_1^+,l_1]=K\,l_0\,,
\end{equation}
where
\begin{equation}
K:=b^\alpha a_\alpha +\bar{a}_{\dot\alpha}\bar{b}^{\dot\alpha}+2
\end{equation}
is a number operator and all other commutators vanish: $[l_1^+,l_0]=0=[l_1,l_0]$.

\subsection{Infinite integer spin fields}

In \cite{Buchbinder:2018yoo}, it was shown that the BRST description of infinite integer spin representations
is carried out by making use of the triplet of vectors
\begin{equation} \label{tr-bos}
|\phi\rangle\,,\quad |\phi_{1}\rangle\,,\quad |\phi_{2}\rangle\,,
\end{equation}
each of which has the form \p{GFState-b}.

The equations of motion for the states \p{tr-bos} are
\begin{equation}
\label{BosEofM}
\begin{array}{l}
\Box|\phi\rangle +(l_1^+{-}\mu)|\phi_{1}\rangle =0\,,
\\ [6pt]
K|\phi_{1}\rangle - (l_1{-}\mu)|\phi\rangle +(l_1^+{-}\mu)|\phi_{2}\rangle
=0\,,
\\ [6pt]
\Box|\phi_{2}\rangle +(l_1{-}\mu)|\phi_{1}\rangle =0\,
\end{array}
\end{equation}
and as it was shown  in \cite{Buchbinder:2018yoo}, these equations of motion follow from the Lagrangian
\begin{eqnarray}
\label{Daction-cs-b}
\mathcal{L}_{\phi}&=&
\langle\bar{\phi}|\Box|\phi\rangle
- \langle\bar{\phi}_1|K|\phi_1\rangle
- \langle\bar{\phi}_2| \Box|\phi_2\rangle
\\ [6pt]
&&{}
+\langle\bar{\phi}|(l_1^+{-}\mu)|\phi_1\rangle
+ \langle\bar{\phi}_1| (l_1{-}\mu)|\phi\rangle
- \langle\bar{\phi}_1|(l_1^+{-}\mu)|\phi_2\rangle
- \langle\bar{\phi}_2|(l_1{-}\mu)|\phi_1\rangle \,.
\nonumber
\end{eqnarray}
The dimensionful nonzero parameter $\mu$ in \p{BosEofM} and \p{Daction-cs-b} defines the value of the four-order Casimir operators
$W^2$ for the massless infinite spin representations of the Poincar\'{e} group \cite{Wigner39,Wigner47,BargWigner}.
The Lagrangian \p{Daction-cs-b} is invariant under transformations
\begin{equation}
\delta|\phi\rangle=(l_1^+{-}\mu)|\lambda\rangle\,,\qquad
\delta|\phi_{1}\rangle=-\Box|\lambda\rangle\,,\qquad
\delta|\phi_{2}\rangle=(l_1{-}\mu)|\lambda\rangle\,,
\label{GTbos-1}
\end{equation}
where the local parameter $|\lambda\rangle$ has the form \p{GFState-b}.

The Lagrangian $\mathcal{L}_{\phi}$ \eqref{Daction-cs-b} for complex
bosonic infinite spin fields $|\phi\rangle$,
$|\phi_{1}\rangle$, $|\phi_{2}\rangle$ will be used bellow as a
bosonic part of the supersymmetric Lagrangian.

\subsection{Infinite half-integer spin fields}

The infinite half-integer spin representation is described in the BRST approach by the triplet \cite{Buchbinder:2020nxn}
\begin{equation} \label{tr-fer}
|\Psi_A\rangle\,,\quad |\Psi_{1}{}_A\rangle\,,\quad |\Psi_{2}{}_A\rangle\,,
\end{equation}
where all vectors have the form \p{GFState}.
Three independent equations of motion have the form
\begin{equation}
\label{Eq1}
\begin{array}{l}
i\!\!\not\!\partial |\Psi\rangle+(l_1^+{-}\mu)|\Psi_{1}\rangle=0\,,
\\ [6pt]
K i\!\!\not\!\partial|\Psi_{1}\rangle-(l_1{-}\mu)|\Psi\rangle+  (l_1^+{-}\mu)|\Psi_{2}\rangle=0\,,
\\ [6pt]
i\!\!\not\!\partial|\Psi_{2}\rangle+(l_1{-}\mu)|\Psi_{1}\rangle=0\,.
\end{array}
\end{equation}
where ${}\not\!\!\partial{}_A{}^B=\partial_m(\gamma^m)_A{}^B$.
Equations \p{Eq1} are Lagrangian equations which can be obtained from the following Lagrangian:
\begin{equation}
\begin{array}{rcl}
\mathcal{L}_{\Psi}&=&
\langle\bar{\Psi}| i\!\!\not\!\partial|\Psi\rangle
- \langle\bar{\Psi}_1| K i\!\!\not\!\partial|\Psi_{1}\rangle
-\langle\bar{\Psi}_2|i\!\!\not\!\partial|\Psi_{2}\rangle
\\ [6pt]
&&{}
+\langle\bar{\Psi}|(l_1^+{-}\mu)|\Psi_{1}\rangle
+\langle\bar{\Psi}_1|(l_1{-}\mu)|\Psi\rangle
-\langle\bar{\Psi}_2|(l_1{-}\mu)|\Psi_{1}\rangle
-\langle\bar{\Psi}_1|(l_1^+{-}\mu)|\Psi_{2}\rangle  \; .
\label{LagrFock}
 \end{array}
\end{equation}
Lagrangian \eqref{LagrFock} is invariant under gauge transformations
\begin{equation}\label{var-BRSTa}
\delta|\Psi\rangle=(l_1^+{-}\mu)|\Pi\rangle \,,\qquad
\delta|\Psi_{1}\rangle=-i\!\!\not\!\partial|\Pi\rangle \,,\qquad
\delta|\Psi_{2}\rangle=(l_1{-}\mu)|\Pi\rangle \,,
\end{equation}
where we omit the external Dirac index, i.e. the local parameter $|\Pi\rangle=|\Pi_A\rangle$ has additional
index $A$ and  is represented in the form \p{GFState}.

In \cite{Buchbinder:2020nxn}, it was noted that Lagrangian
\eqref{LagrFock} contains twice as many degrees of freedom as
Metsaev's Lagrangian \cite{Metsaev:2017ytk}. Now we split Lagrangian
\eqref{LagrFock} into two parts, each of which will describe the
fermionic field of a infinite spin. For this purpose we
decompose the Dirac spinors \p{tr-fer} in terms of the sum of two
Weyl spinors in the following form:
\begin{equation} \label{dec-Weyl-2a}
|\Psi_A\rangle =\left(\!
\begin{array}{c}
|\psi_\alpha\rangle \\ [5pt]
|\bar\chi^{\dot\alpha}\rangle \\
\end{array}
\!\right),\qquad
|\Psi_{\,1 A}\rangle =\left(\!
\begin{array}{c}
|\chi_1{}_\alpha\rangle \\ [5pt]
|\bar\psi_1^{\dot\alpha}\rangle \\
\end{array}
\!\right),\qquad
|\Psi_{\,2 A}\rangle =\left(\!
\begin{array}{c}
|\psi_2{}_\alpha\rangle \\ [5pt]
|\bar\chi_2^{\dot\alpha}\rangle \\
\end{array}
\!\right),
\end{equation}
and
\begin{equation} \label{dec-Weyl-ca}
\langle\bar\Psi^{A}| =\left( \langle\chi^{\alpha}|, \langle\bar\psi_{\dot\alpha}| \right),\qquad
\langle\bar\Psi_1^{A}| =\left( \langle\psi_1^{\alpha}|, \langle\bar\chi_1{}_{\dot\alpha}| \right),\qquad
\langle\bar\Psi_2^{A}| =\left( \langle\chi_2^{\alpha}|, \langle\bar\psi_2{}_{\dot\alpha}| \right),
\end{equation}
where
\begin{equation} \label{conj-Weyl-comp}
\begin{array}{ll}
\left(|\psi_\alpha\rangle\right)^\dagger= \langle\bar\psi_{\dot\alpha}| \,,\quad&
\left(|\bar\chi^{\dot\alpha}\rangle\right)^\dagger= \langle\chi^{\alpha}|\,,\\ [5pt]
\left(|\psi_1{}_\alpha\rangle\right)^\dagger= \langle\bar\psi_{1\dot\alpha}| \,,\quad&
\left(|\bar\chi_1^{\dot\alpha}\rangle\right)^\dagger= \langle\chi_1^{\alpha}|\,,\\ [5pt]
\left(|\psi_2{}_\alpha\rangle\right)^\dagger= \langle\bar\psi_{2\dot\alpha}| \,,\quad&
\left(|\bar\chi_2^{\dot\alpha}\rangle\right)^\dagger= \langle\chi_2^{\alpha}|\,.
\end{array}
\end{equation}

In terms of the Weyl spinors Lagrangian \p{LagrFock} takes the form
\begin{equation} \label{LagrFockw}
\mathcal{L}_{\Psi}
=
\mathcal{L}_{\psi}+\mathcal{L}_{\chi}
\end{equation}
where the chiral parts of $\mathcal{L}_{\Psi}$ are
\begin{eqnarray}
\label{LagrFockwpsi}
\mathcal{L}_{\psi} &=&
\langle\bar{\psi}_{\dot{\alpha}}|i\partial^{\dot\alpha\alpha}|\psi_\alpha\rangle
 -  \langle{\psi}_1^{\alpha}|Ki\partial_{\alpha\dot\alpha}|\bar\psi_1^{\dot\alpha}\rangle
 -  \langle\bar{\psi}_{2\dot{\alpha}}|i\partial^{\dot\alpha\alpha}|\psi_{2\alpha}\rangle
\\[5pt]
&&{}
+  \langle\bar{\psi}_{\dot\alpha}|(l^+_1{-}\mu)|\bar{\psi}_1^{\dot\alpha}\rangle
+  \langle\psi_1^\alpha|(l_1{-}\mu)|\psi_{\alpha}\rangle
+  \langle\bar{\psi}_{\dot\alpha}|(l^+_1{-}\mu)|\bar{\psi}_1^{\dot\alpha}\rangle
+  \langle\psi_1^\alpha|(l_1{-}\mu)|\psi_{\alpha}\rangle \,, \nonumber \\ [7pt]
\label{LagrFockwchi}
\mathcal{L}_{\chi} &=&
\langle\bar{\chi}_{\dot{\alpha}}|i\partial^{\dot\alpha\alpha}|\chi_\alpha\rangle
-  \langle{\chi}_1^{\alpha}|Ki\partial_{\alpha\dot\alpha}|\bar\chi_1^{\dot\alpha}\rangle
-  \langle\bar{\chi}_{2\dot{\alpha}}|i\partial^{\dot\alpha\alpha}|\chi_{2\alpha}\rangle
\\[5pt]
&&{}
-  \langle\bar{\chi}_{\dot\alpha}|(l^+_1{-}\mu)|\bar{\chi}_1^{\dot\alpha}\rangle
-  \langle\chi_1^\alpha|(l_1{-}\mu)|\chi_{\alpha}\rangle
-  \langle\bar{\chi}_{\dot\alpha}|(l^+_1{-}\mu)|\bar{\chi}_1^{\dot\alpha}\rangle
-  \langle\chi_1^\alpha|(l_1{-}\mu)|\chi_{\alpha}\rangle \,.
\nonumber
\end{eqnarray}
Gauge transformations \eqref{var-BRSTa} in terms of the Weyl spinors are
\begin{equation}\label{var-BRSTa-1}
\delta|\psi_\alpha\rangle=(l_1^+-\mu)|\pi_\alpha\rangle\,,\qquad
\delta|\bar{\psi}_1^{\dot{\alpha}}\rangle=-i\partial^{\dot\alpha\beta}|\pi_\beta\rangle \,,\qquad
\delta|\psi_{2\alpha}\rangle=(l_1-\mu)|\pi_\alpha\rangle\,,
\end{equation}
\begin{equation}\label{var-BRSTa-2}
\delta|\bar{\chi}^{\dot{\alpha}}\rangle=(l_1^+-\mu)|\bar\rho^{\dot\alpha}\rangle\,,\qquad
\delta|\chi_{1\alpha}\rangle=-i\partial_{\alpha\dot\beta}|\bar\rho^{\dot\beta}\rangle\,,\qquad
\delta|\bar{\chi}_2^{\dot{\alpha}}\rangle=(l_1-\mu)|\bar\rho^{\dot\alpha}\rangle\,,
\end{equation}
where $|\lambda_\alpha\rangle$ and $|\bar\rho^{\dot\alpha}\rangle$ are the Weyl components of the Dirac ket-vector of the gauge parameter:
\begin{equation} \label{dec-Weyl-2b}
|\Pi_A\rangle =\left(\!
\begin{array}{c}
|\pi_\alpha\rangle \\ [5pt]
|\bar\rho^{\dot\alpha}\rangle \\
\end{array}
\!\right).
\end{equation}

We see that the Lagrangian \p{LagrFock} and gauge transformations \eqref{var-BRSTa} split into two independent
parts, each of which contains Weyl
halves of the Dirac spinors.
Each of the chiral parts $\mathcal{L}_{\psi}$ and $\mathcal{L}_{\chi}$ of the Lagrangian $\mathcal{L}_{\Psi}$
describes the infinite spin multiplet with half of the fermion states of the
infinite spin multiplet defined by the Lagrangian $\mathcal{L}_{\Psi}$, which was proposed in
\cite{Buchbinder:2020nxn}.

In what follows, we will consider the chiral part $\mathcal{L}_{\psi}$ \eqref{LagrFockwpsi} as
constituent part of a supersymmetric system.

\section{Superfield description}\label{sec:SField}

Let us consider the system with the Lagrangian
\begin{equation} \label{L-susy}
\mathcal{L}=\mathcal{L}_{\phi} \ -\ \frac12\,\mathcal{L}_\psi\,,
\end{equation}

This Lagrangian
is invariant under the following supersymmetry variations
\begin{equation}
\begin{array}{lllll}
&\delta|\phi\rangle=\epsilon^\beta |\psi_\beta\rangle\,,
&&&\delta\langle\bar{\phi}|= \langle\bar{\psi}_{\dot\beta}|\,\bar{\epsilon}^{\dot\beta}\,,
\\ [5pt]
&\delta |\psi_\alpha\rangle =2i\bar{\epsilon}^{\dot\beta}\partial_{\alpha\dot\beta}|\phi\rangle\,,
&&&\delta \langle\bar{\psi}_{\dot\alpha}| =-2i\partial_{\beta\dot\alpha}\langle\bar{\phi}|\,\epsilon^\beta\,,
\\[0.9em]
&\delta|\phi_1\rangle= i\,\epsilon^\beta\partial_{\beta\dot\alpha}|\bar{\psi}_1^{\dot\alpha}\rangle\,,
&&&\delta\langle\bar{\phi}_1|= -i\,\partial_{\alpha\dot\beta}\langle\psi_1^\alpha|\,\bar{\epsilon}^{\dot\beta}\,,
\\ [5pt]
&\delta |\bar{\psi}_1^{\dot\alpha}\rangle = 2 \,{\bar\epsilon}^{\dot\alpha}|\phi_1\rangle\,,
&&&\delta\langle\psi_1^\alpha|=2 \,\langle\bar{\phi}_1|\,\epsilon^\alpha \,,
\\[0.9em]
&\delta|\phi_2\rangle= \epsilon^\beta |\psi_{2\beta}\rangle\,,
&&& \delta\langle\bar{\phi}_2|= \langle\bar{\psi}_{2\dot\beta}| \,\bar{\epsilon}^{\dot\beta} \,,
\\ [5pt]
&\delta |\psi_{2\alpha}\rangle =2i\bar{\epsilon}^{\dot\beta}\partial_{\alpha\dot\beta}|\phi_2\rangle\,,
&&& \delta \langle\bar{\psi}_{2\dot\alpha}| =-2i\partial_{\beta\dot\alpha}\langle\bar{\phi}_2|\,\epsilon^\beta \,.
\end{array}
\label{susytr1}
\end{equation}
One can show that the algebra of these transformations is not closed,
which is typical for supersymmetry transformations without
auxiliary fields (on-shell supersymmetry).

Now we turn attention that the supersymmetry transformations \eqref{susytr1} allow one to
treat the above vectors as chiral multiplets of
$\mathcal{N}=1$ supersymmetry. More precisely, the pairs of the
ket-vectors
\begin{equation} \label{ket-chir-on}
\Big(|\phi\rangle\,, |\psi_\alpha\rangle\Big)\,,
\qquad
\Big(|{\phi}_2\rangle\,, |\psi_{2\alpha}\rangle\Big)
\end{equation}
form chiral multiplets, while the other pair of the ket-states forms antichiral multiplet
\begin{equation} \label{ket-achir-on}
\Big(|\bar\psi_1^{\dot\alpha}\rangle\,, |{\phi}_1\rangle\Big)\,.
\end{equation}
Conjugated bra-states have opposite chirality: pairs of bra-states
\begin{equation} \label{bra-achir-on}
\Big(\langle\bar{\phi}|\,, \langle\bar\psi_{\dot\alpha}|\Big)\,,
\qquad \Big(\langle\bar{\phi}_2|\,, \langle\bar\psi_{2\dot\alpha}|\Big)
\end{equation}
are antichiral while a pair of bra-states
\begin{equation} \label{bar-chir-on}
\Big(\langle\psi_1^{\alpha}|\,, \langle\bar{\phi}_1|\Big)
\end{equation}
is chiral.
For the off-shell supersymmetric description, we will add auxiliary components and construct the
corresponding (anti)chiral superfields.

Off-shell $\mathcal{N}=1$ supertranslations are realized in $\mathcal{N}=1$ superspace with
supercoordinates $(x^m,\theta^\alpha,\bar\theta^{\dot\alpha})$ as follows:
\begin{equation} \label{susy-tr}
\delta x^m = i\,\theta\sigma^m\bar\epsilon \ - \ i\,\epsilon\sigma^m\bar\theta\,,\qquad \delta\theta^{\alpha}=\epsilon^{\alpha}\,,\qquad
\delta\bar\theta^{\dot\alpha}=\bar\epsilon^{\dot\alpha}\,.
\end{equation}
The corresponding supersymmetry generators and covariant derivatives have the form
\begin{equation}
\label{susy-gen}
Q_{\alpha} = \frac{\partial}{\partial \theta^{\alpha}}-i\partial_{\alpha\dot\beta}\bar\theta^{\dot\beta}\,,
\quad
\bar Q_{\dot\alpha} = -\displaystyle\frac{\partial}{\partial \bar\theta^{\dot\alpha}}+i\theta^{\beta}\partial_{\beta\dot\alpha}\,,
\end{equation}
\begin{equation}
\label{susy-der}
D_{\alpha} = \displaystyle\frac{\partial}{\partial \theta^{\alpha}}+i\partial_{\alpha\dot\beta}\bar\theta^{\dot\beta}\,,
\quad
\bar D_{\dot\alpha} = -\displaystyle\frac{\partial}{\partial \bar\theta^{\dot\alpha}}-i\theta^{\beta}\partial_{\beta\dot\alpha}\,,
\end{equation}

Now we add three additional fields
\begin{equation}
\label{add-fields}
|f(x_L)\rangle\,, \qquad |e(x_R)\rangle\,, \qquad |g(x_L)\rangle
\end{equation}
to three pairs of states \p{ket-chir-on}, \p{ket-achir-on} to form three
(anti)chiral superfield ket-vectors
\begin{equation}
\label{susy-fields-ch}
\begin{array}{rcl}
|\Phi(x_L,\theta)\rangle &=& |\phi(x_L)\rangle \ + \ \theta^\alpha|\psi_\alpha(x_L)\rangle \ + \ \theta^\alpha\theta_\alpha |f(x_L)\rangle\,,
\\[6pt]
|\bar S_1(x_R,\bar\theta)\rangle &=& |e(x_R)\rangle \ + \ \bar\theta_{\dot\alpha}|\bar\psi_1^{\dot\alpha}(x_R)\rangle \ + \
\bar\theta_{\dot\alpha}\bar\theta^{\dot\alpha}  |\phi_1(x_R)\rangle\,,
\\[6pt]
|S_2(x_L,\theta)\rangle  &=& |\phi_2(x_L)\rangle \ + \ \theta^\alpha|\psi_{2\alpha}(x_L)\rangle \ + \ \theta^\alpha\theta_\alpha |g(x_L)\rangle\,,
\end{array}
\end{equation}
where
\begin{equation} \label{x-L}
x_L^m = x^m+i\,\theta\sigma^m\bar\theta\,,
\qquad
x_R^m = x^m-i\,\theta\sigma^m\bar\theta\,.
\end{equation}

Conjugated states \p{bra-achir-on}, \p{bar-chir-on} form three (anti)chiral superfield bra-vectors
\begin{equation}
\label{susy-fields-ach}
\begin{array}{rcl}
\langle\bar{\Phi}(x_R,\bar\theta)| &=& \langle \bar\phi(x_R)| \ + \ \langle \bar\psi_{\dot\alpha}(x_R)|\bar\theta^{\dot\alpha} \ + \
\langle \bar f(x_R)|\bar\theta_{\dot\alpha}\bar\theta^{\dot\alpha}\,,
\\[6pt]
\langle S_1(x_L,\bar\theta)| &=& \langle \bar e(x_L)| \ + \ \langle \psi^{\alpha}_1(x_L)|\theta_{\alpha} \ + \
\langle \bar\phi_1(x_L)|\theta^\alpha\theta_\alpha \,,
\\[6pt]
\langle \bar S_2(x_R,\bar\theta)|  &=& \langle \bar\phi_2(x_R)| \ + \ \langle \bar\psi_{2{\dot\alpha}}(x_R)|\bar\theta^{\dot\alpha} \ + \
\langle \bar g(x_R)|\bar\theta_{\dot\alpha}\bar\theta^{\dot\alpha} \,.
\end{array}
\end{equation}

Taking into account the general supertranslation formula $\delta F(x,\theta,\bar\theta)=\left(\epsilon^\alpha Q_\alpha +\bar\epsilon_{\dot\alpha} \bar Q^{\dot\alpha}\right)F$,
the superfield representations \p{susy-fields-ch} lead to the following variations of the component states:
\begin{equation}
\label{tr-c}
\begin{array}{lll}
\delta|\phi\rangle=\epsilon^\beta |\psi_\beta\rangle\,, \quad&
\delta |\psi_\alpha\rangle =2i\bar{\epsilon}^{\dot\beta}\partial_{\alpha\dot\beta}|\phi\rangle+2\epsilon_\alpha |f\rangle\,, \quad&
\delta|f\rangle=i\bar\epsilon_{\dot\beta}\partial^{\dot\beta\alpha} |\psi_\alpha\rangle\,,
\\ [6pt]
\delta|e\rangle=\epsilon_{\dot\beta} |\bar\psi_1^{\dot\beta}\rangle\,, \quad&
\delta |\bar{\psi}_1^{\dot\alpha}\rangle = 2i\epsilon_\beta \partial^{\beta\dot\alpha}|e\rangle
+2{\bar\epsilon}^{\dot\alpha}|\phi_1\rangle\,,
\quad&
\delta|\phi_1\rangle= i\epsilon^\beta\partial_{\beta\dot\alpha}|\bar{\psi}_1^{\dot\alpha}\rangle\,,
\\ [6pt]
\delta|\phi_2\rangle=\epsilon^\beta |\psi_2{}_\beta\rangle\,, \quad&
\delta |\psi_2{}_\alpha\rangle =2i\bar{\epsilon}^{\dot\beta}\partial_{\alpha\dot\beta}|\phi_2\rangle+2\epsilon_\alpha |g\rangle\,, \quad&
\delta|g\rangle=i\bar\epsilon_{\dot\beta}\partial^{\dot\beta\alpha} |\psi_2{}_\alpha\rangle\,,
\end{array}
\end{equation}
\begin{equation}
\label{tr-ac}
\begin{array}{lll}
\delta\langle\bar{\phi}|= \langle\bar{\psi}_{\dot\beta}|\,\bar{\epsilon}^{\dot\beta}\,, \quad&
\delta \langle\bar{\psi}_{\dot\alpha}| =-2i\partial_{\beta\dot\alpha}\langle\bar{\phi}|\,\epsilon^\beta
+2\langle\bar f|\,\bar\epsilon_{\dot\alpha}\,, \quad&
\delta\langle\bar f|=i\partial^{\dot\alpha\beta} \langle\bar{\psi}_{\dot\alpha}| \epsilon_{\beta}\,,
\\ [6pt]
\delta\langle\bar e|= \langle\psi_1^{\beta}|\,\epsilon_{\beta}\,, \quad&
\delta\langle\psi_1^\alpha|=-2i\partial^{\dot\beta\alpha}\langle\bar{e}|\,\bar\epsilon_{\dot\beta}
+2 \,\langle\bar{\phi}_1|\,\epsilon^\alpha \,,
\quad&
\delta\langle\bar{\phi}_1|= -i\,\partial_{\alpha\dot\beta}\langle\psi_1^\alpha|\,\bar{\epsilon}^{\dot\beta}\,,
\\ [6pt]
\delta\langle\bar{\phi}_2|= \langle\bar{\psi}_2{}_{\dot\beta}|\,\bar{\epsilon}^{\dot\beta}\,,
\quad&
\delta \langle\bar{\psi}_2{}_{\dot\alpha}| =-2i\partial_{\beta\dot\alpha}\langle\bar{\phi}_2|\,\epsilon^\beta
+2\langle\bar g|\,\bar\epsilon_{\dot\alpha}\,,
\quad&
\delta\langle\bar g|=i\partial^{\dot\alpha\beta} \langle\bar{\psi}_2{}_{\dot\alpha}| \epsilon_{\beta}\,.
\end{array}
\end{equation}
As it will be shown later, the field $|e\rangle$ can be removed by
means of its gauge transformation \eqref{eh} and then the states
$|f\rangle$, $|g\rangle$ vanish due to their equations of motion:
$|f\rangle =0$, $|g\rangle =0$. After this, supersymmetry
transformations \p{tr-c}, \p{tr-ac} pass to the on-shell invariance
\p{susytr1} of the Lagrangian \p{L-susy}.

\section{Superfield Lagrangian}\label{sec:SFieldLagr}

Let us consider the superfield Lagrangian
\begin{eqnarray} \label{L-susy-f}
\!\!\!\mathcal{L}_{super}\!\! &\!=\!& \!\! \int d^2\theta d^2\bar\theta \,
\Big(
\langle\bar\Phi| \Phi\rangle
-\langle S_1|K |\bar S_1\rangle
-\langle\bar S_2|S_2\rangle
\Big)
\\ [5pt]
&& +\int d^2\theta \,
\Big(
\langle S_1|(l_1-\mu)| \Phi\rangle
-\langle S_1|(l^+_1-\mu)| S_2\rangle
\Big)
\nonumber
\\ [5pt]
&&
+\int d^2\bar\theta \,
\Big(
\langle\bar \Phi|(l_1^+-\mu)|\bar S_1\rangle
-\langle\bar S_2|(l_1-\mu)|\bar S_1\rangle
\Big),
\nonumber
\end{eqnarray}
where
$$
d^2\theta=-\frac14\, d\theta^\alpha d\theta^\beta\epsilon_{\alpha\beta}\,,\quad
d^2\bar\theta=-\frac14\, d\bar\theta_{\dot\alpha} d\bar\theta_{\dot\beta}\epsilon^{\dot\alpha\dot\beta}\,,\qquad
\int \theta^\alpha \theta_\alpha\,d^2\theta=1\,,\quad
\int \theta_{\dot\alpha}\theta^{\dot\alpha}\,d^2\bar\theta=1\,.
$$

The Lagrangian \eqref{L-susy-f} is invariant under the gauge
transformations
\begin{align}
&\delta |\Phi\rangle=(l_1^+-\mu)|\Lambda\rangle,
&&\delta |\bar S_1\rangle=\tfrac{1}{4}D^2|\Lambda\rangle
\label{SGT}
&&\delta|S_2\rangle=(l_1-\mu)|\Lambda\rangle,
\end{align}
where $|\Lambda\rangle$ is the chiral superfield
\begin{eqnarray} 
|\Lambda(x_L,\theta)\rangle &=&
|\lambda(x_L)\rangle \ + \ \theta^\alpha|\pi_\alpha(x_L)\rangle \ + \ \theta^\alpha\theta_\alpha |\varepsilon(x_L)\rangle
\,.
\end{eqnarray}
The gauge transformation \eqref{SGT} in the component form gives \eqref{GTbos-1}, \eqref{var-BRSTa-1} and for the additional fields
\begin{align}\label{eh}
&\delta |f\rangle=(l_1^+-\mu) |\varepsilon\rangle\,,
&&\delta |e\rangle=-|\varepsilon\rangle\,,
&&\delta |g\rangle=(l_1-\mu)|\varepsilon\rangle\,.
\end{align}

After Grassmannian integration the Lagrangian \p{L-susy-f} acquires the following component form
\begin{equation} \label{L-susy-f-comp}
\mathcal{L}_{\text{super}} \ = \ \mathcal{L}_{\phi} \ -\ \frac12\,\mathcal{L}_\psi
+\mathcal{L}_{\text{add}} \,,
\end{equation}
where the first two terms
$\mathcal{L}_{\phi}-\frac12\,\mathcal{L}_\psi$ give exactly the
Lagrangian \p{L-susy} while the last term has the form
\begin{eqnarray}
\mathcal{L}_{\text{add}}
&=&
\langle\bar f| f\rangle
-\langle\bar e|K\Box |e\rangle
-\langle\bar g|g\rangle +
\label{Laux}
\\ [5pt]
&&
+ \langle\bar e|(l_1-\mu)| f\rangle + \langle\bar f|(l^+_1-\mu)| e\rangle
-\langle\bar e|(l^+_1-\mu)| g\rangle -\langle\bar g|(l_1-\mu)| e\rangle \,.
\nonumber
\end{eqnarray}

We see that in \p{L-susy-f-comp} the Lagrangian
$\mathcal{L}_{\text{add}}$ for the additional fields is uncoupled
from dynamical Lagrangian \p{L-susy}. Moreover, we can remove the
state $|e\rangle$ by the gauge-fixing condition $|e\rangle=0$ for
the gauge symmetry \p{eh}. Then the additional states $|f\rangle$,
$|g\rangle$ are eliminated with the help of their equations of
motion $|f\rangle=0$, $|g\rangle=0$. As a result of this, we obtain
Lagrangian \p{L-susy} after eliminating some of the auxiliary and
gauge states in the system described by Lagrangian
\p{L-susy-f-comp}, which is equivalent to the superfield Lagrangian
\p{L-susy-f}.

Thus we have shown that Lagrangian \eqref{L-susy-f} indeed is the superfield Lagrangian for the
supersymmetric infinite spin system.

\section{Summary and outlook}

Let us summarize the results. We have presented the superfield
formulation of the $\mathcal{N}=1$ supersymmetric generalization of
the $4D$ infinite spin theory. To construct a supersymmetric theory
of infinite spin, we have used our Lagrangians for bosonic
\cite{Buchbinder:2018yoo} and fermionic \cite{Buchbinder:2020nxn}
infinite spin fields constructed in terms of spin-tensor fields with
dotted and undotted indices on the basis of the BRST approach. At
the same time, we showed that, in contrast to
\cite{Buchbinder:2020nxn}, one can consider a fermionic infinite
spin multiplet whose component fields cooperate into a generating
field with external Weyl index. This field of infinite spin
describes half as many degrees of freedom as the Dirac field of
infinite spin from \cite{Buchbinder:2020nxn}. First, we consider the
sum of Lagrangians \p{Daction-cs-b} and \p{LagrFockwpsi},
respectively,  for the triplet of complex bosonic fields and for the
triplet of  fermionic Weyl fields. Such content of the fields is due
to the desire to compose a chiral superfield. Then we first find
supersymmetry transformation \p{susytr1} for such a sum of the
Lagrangians in component form and then we add several auxiliary
fields \p{add-fields} and construct the superfield Lagrangian
\p{L-susy-f} for infinite spin in terms of three pairs
chiral and anti-chiral  superfields \p{susy-fields-ch}. The
important point here is that we need supersymmetric generalization
\p{SGT} of the BRST-like gauge invariance, so that it preserves the
chirality, or antichirality, of the corresponding superfield states.
Thus, in the present paper, we have obtained the superfield
Lagrangian formulation for the $4D, \mathcal{N}=1$ supersymmetric
infinite spin theory.

As a continuation of this research, it would be interesting to
construct supersymmetric infinite spin systems with extended
$\mathcal{N}>1$ supersymmetry. In addition, within the framework of
the BRST-like approach we plan to construct field
descriptions of infinite spin systems in higher space-time
dimensions. In particular, we plan to construct such a description
in six dimensions based on our recent works
\cite{PFIP:2021,PFI:2021}.

\section*{Acknowledgments}
The authors are grateful to M. Najafizadeh for useful comment.  The
work of I.L.B. and S.A.F. was supported by the Russian Science
Foundation, project no. 21-12-00129. The work of A.P.I. and V.A.K.
was supported by the Ministry of Education of the Russian
Federation, project FEWF-2020-0003.

\end{document}